\documentclass{ifacconf}

\usepackage{graphicx}      
\usepackage{xcolor}
\usepackage{url}
\usepackage{natbib}        
\usepackage{amsmath,amssymb,amsfonts,mathtools}
\let\classAND\AND
\let\AND\relax
\usepackage{algorithm,algorithmic}

\let\AND\classAND
\AtBeginEnvironment{algorithmic}{\let\AND\algoAND}

\counterwithin*{section}{part}
\usepackage{cleveref}
\crefname{equation}{}{}
\crefrangelabelformat{equation}{\textup{(#3#1#4-#5\crefstripprefix{#1}{#2}#6)}}

\begin{document}
\begin{frontmatter}
\title{Predictive stability filters for nonlinear dynamical systems affected by disturbances} 
\author[ICS,FA]{Alexandre Didier}
\author[ICS,FA]{Andrea Zanelli}
\author[ICS]{Kim P. Wabersich}
\author[ICS]{Melanie N. Zeilinger}
\address[ICS]{Institute for Dynamic Systems and Control, \\ ETH Zürich, 8092 Zürich, Switzerland \\(e-mail: \texttt{\{adidier,zanellia,wkim,mzeilinger\}@ethz.ch}). }   
\address[FA]{Contributed equally to this paper.}
\begin{abstract}                
Predictive safety filters provide a way of projecting potentially unsafe inputs, proposed, e.g. by a human or learning-based controller, onto the set of inputs that guarantee recursive state and input constraint satisfaction by leveraging model predictive control techniques.
In this paper, we extend this framework such that in addition, robust asymptotic stability of the closed-loop system can be guaranteed by enforcing a decrease of an implicit Lyapunov function which is constructed using a predicted system trajectory.  
Differently from previous results, we show robust asymptotic stability with respect to a predefined disturbance set on an extended state consisting of the system state and a warmstart input sequence. 
The proposed strategy is applied to an automotive lane keeping example in simulation.
\end{abstract}
\begin{keyword}
Robust Model Predictive Control, Stability and Recursive Feasibility, Automotive
\end{keyword}
\end{frontmatter}
\section{Introduction}
Safety filters provide a modular framework, which allow certifying safety of dynamical systems in terms of constraint satisfaction, see, e.g. \cite{Wabersich2023} for an overview. In order to certify safety for complex systems affected, e.g., by nonlinearities, high-dimensionality, constraints and 
unknown disturbances, model predictive control concepts (MPC), see, e.g., \cite{Rawlings2017}, can be utilized in order to design a predictive safety filter, as proposed in \cite{Wabersich2018, Wabersich2021}.
This filter ideally interferes as little as possible with the primary control objective. However, as this objective can potentially be unknown, the filter instead minimizes the distance of a proposed input and the input applied to the system.
In this manner, the safety filter guarantees safety while remaining agnostic to the primary performance objective. 
In particular, the proposed input could come from, e.g., a human operating a safety critical system, such as an automotive, aerospace or biomedical system, or a learning algorithm which is learning a policy or a model. 
\par
While constraint satisfaction is the minimal requirement for safety, it can often be beneficial for the closed-loop performance to enforce stronger closed-loop system properties through the safety filter. A prominent application domain is given by automotive systems where one might additionally want to enforce stability properties during non-emergency related interventions, which provide more comfort or increased performance. 
As a motivating example, consider a predictive safety filter applied to  
a vehicle, controlled by a human driver on a high-way. A safety filter that enforces constraint satisfaction would prevent the vehicle from crashing when unsafe inputs are applied due to, e.g., a distracted driver. However, it is additionally desirable to guarantee that the vehicle eventually converges to a high-way lane after avoiding a crash or incurring disturbances
and that the vehicle does not sustain large driver-induced oscillations (DIO), due to human delay of response. In general, common safety filter methods can facilitate or cause unwanted oscillations and/or convergence to spurious equilibria, e.g., when transitioning from emergency interventions in airplanes to a stable flight. 
\par
In this work, we focus on the additional enforcement of robust asymptotic stability under application of a predictive safety filter.
The stability result is achieved by including a decrease constraint of a Lyapunov function, defined in the set of states and feasible input trajectories and constructed implicitly through stage and terminal costs.
\subsection{Related work and contribution}\label{sec:lit}
The presented work builds on predictive safety filters and is meant to provide an extension that guarantees not only constraint satisfaction, but stability of the underlying closed-loop dynamics. In particular, \cite{Wabersich2018} introduce the concept of predictive safety filters for linear systems with additive disturbances as an alternative to invariance-based safety filters such as, e.g., \cite{Akametalu2014} and \cite{Fisac2019}, with an explicit specification of \textit{safe sets}. The concept is then extended for nonlinear uncertain systems in \cite{Wabersich2021}. 
An overview of different safety filter methods can be found in \cite{Wabersich2023}.
\par
In order to obtain predictive filters with stabilizing properties, the proposed approach relies on the enforcement of the decrease of a Lyapunov function through a so-called \textit{Lyapunov constraint}. Several works exist in the literature that rely on a similar idea in order to guarantee stability of the resulting closed-loop system. 
\par
The main idea and analysis provided in this work draws from ideas in \textit{suboptimal MPC} such as the ones in, e.g., \cite{Scokaert1999}, \cite{Pannocchia2011}, \cite{Zeilinger2014}, \cite{Allan2017} and \cite{Mcallister2023}. In particular, the analysis in \cite{Allan2017} provides suitable tools 
for analyzing the proposed approach in the sense that it introduces the notion of \textit{augmented state-warmstart dynamics} and \textit{difference inclusions} that allow one to analyze the evolution of the system for any feasible solution and, in our case, any value of the proposed input. Our work differs from \cite{Allan2017}, and all the suboptimal MPC literature covered by it, in the sense that \textit{i)} we do not optimize for the same cost that defines the Lyapunov function, but rather for an arbitrary cost \textit{ii)} we leverage so-called robust by design strategies rather than relying on inherent robustness properties.
\par
In a similar manner, the field of Lyapunov-based MPC, see, e.g., \cite{Mhaskar2006} and \cite{Heidarinejad2012}, exploits MPC controllers that leverage a \textit{pre-existing} Lyapunov function in order to guarantee stability of the closed-loop dynamics. To this end, a constraint that enforces a decrease of the Lyapunov function is incorporated in the underlying parametric optimization programs.
Similarly, safety filters based on control barrier functions, such as, e.g., \cite{Agrawal2017}, \cite{Greeff2021} and \cite{Didier2023}, enforce a decrease of an explicit control barrier function in order to achieve stability of a safe set of states, possibly in addition to a Lyapunov function decrease, in the constraints.
With respect to these methods, we \textit{i)} regard uncertain systems and provide robust constraint satisfaction guarantees \textit{ii)} do not require that a Lyapunov function is computed explicitly, but rather implicitly through an MPC-like design.
\par
Finally, the work in \cite{Soloperto2020} proposes an algorithm for \textit{``augmentation''} of MPC with active learning and the algorithm proposed in the present paper can be seen as a special case 
thereof. With respect to \cite{Soloperto2020}, we propose a rigorous stability analysis of the resulting \textit{augmented} dynamics.
\par
The paper is structured as follows. In Section \ref{sec:intro} we introduce the problem setting and in Section~\ref{sec:stab_filter}, we propose the stability filter formulation and the corresponding online algorithm. In Section \ref{sec:main_result} we state the main theoretical result of the paper, i.e., a proof of robust asymptotic stability of the augmented state-warmstart dynamics. A suitable MPC method for the stability filter design is discussed in Section \ref{sec:methods} and is applied in simulation to an industry-relevant example, i.e., the stability of a vehicle's dynamics. 

\subsection{Notation}
Throughout the paper we will denote the Euclidean norm of a vector by $\| \cdot \|$. 
All vectors are column vectors and we denote the concatenation of two vectors 
by $(x,y)\vcentcolon=\begin{bmatrix}x^{\top} & y^{\top}\end{bmatrix}^{\top}$.
A sequence of elements $x_i$ with $i\geq 0$ is denoted in bold, i.e. $\mathbf{x} := \{x_0, x_1, \dots, x_N\}$, where the length $N$ of the sequence can be inferred from context, and the sequence norm $\| \mathbf{x} \| := \sup_{i} \| x_i \|$.
The Euclidean ball of radius $r$ centered at $x$ is denoted as $\mathcal{B}(x,r) \vcentcolon=\{ y \,\mid \,\| x - y \| \leq r\}$.
We denote the Minkowski sum of two sets by $\oplus$, i.e., $A \oplus B = \{ a + b \vcentcolon a \in A, b \in B \}$.
Finally, we denote the identity matrix by $\mathbb{I}$.

\section{Preliminaries}\label{sec:intro}
Consider the discrete-time dynamical system, 
\begin{equation}\label{eq:system}
    x^+ = f(x,u,w),
\end{equation}
where $x \in \mathbb{R}^{n_x}$, $u \in \mathbb{R}^{n_u}$ denote the
state and input of the system, respectively, and $w \in \mathcal{W} \subset \mathbb{R}^{n_w}$ is a disturbance. We assume that the origin is an equilibrium point of the system $x^+=f(x,u,0)$ and that $\mathcal{W}$ is compact and contains the origin.
The system is subject to compact state and input constraints of the form $(x,u)\in\mathcal{Z}$, 
which contain the origin and are required to be satisfied at every time step of operation.
\par
In this paper, we consider the problem of filtering inputs of some arbitrary controller online, such that safety in terms of constraint satisfaction is guaranteed. Specifically, we consider predictive safety filters, which were introduced in \citet{Wabersich2018,Wabersich2021}, of the form:
\begin{subequations}\label{eq:safety_filter}
\begin{align}
 \underset{\begin{subarray}{c}
    z_0, \dots, z_N \\
    v_0, \dots, v_{N-1}
 \end{subarray}}{\min} &  J(\mathbf{z},\mathbf{v},p)\\
 \text{s.t. } \;\;\;  & z_{0} = x, \label{eq:constr1sf}\\
        	  & z_{i+1} = f(z_i,v_i, 0), \, i=0,\dots,N-1,\\
          	& (z_i, v_i) \in \Omega_i, \quad \quad \,\,\,\, i=0,\dots,N-1,\\
         	& z_N \in \mathcal{Z}_{f}.
 \end{align}
 \end{subequations}
At every time step, a nominal trajectory of states $\mathbf{z} \in \mathbb{R}^{n_x \cdot (N+1)}$, initialized at the current state $x$ in \eqref{eq:constr1sf} and inputs $\mathbf{v} \in \mathbb{R}^{n_u {\cdot} N}$ are predicted up to some horizon $N \in \mathbb{N}$. The variable $p \in \mathbb{R}^{n_p}$ is some parameter which can change across 
instances, e.g., an input $u_L \in \mathbb{R}^{n_u}$ proposed by a learning-based or human controller as detailed in \cite{Wabersich2018}. The function
$J \!\vcentcolon\! \mathbb{R}^{n_x \!\cdot\! (N+1)} \times \mathbb{R}^{n_u \cdot N} \times \mathbb{R}^{n_p} \rightarrow \mathbb{R}$ 
denotes the objective function which can be chosen, e.g., to be a projection cost associated with the learning input, i.e., $J(\mathbf{z},\mathbf{v},p) = \| v_0 - u_{L} \|^2$.
The sets $\Omega_i$ ,which contain the predicted trajectory, are \textit{tightened} state and input constraints, which are designed offline together with the terminal set $\mathcal{Z}_f$ in order to ensure robust constraint satisfaction in closed-loop operation. 
\par
The formal requirements on the tightened constraint sets $\Omega_i$ and the terminal set $\mathcal{Z}_f$ are provided in Assumption~\ref{assum:improved_candidate}. A design methods which satisfies these formal requirements for linear systems is reviewed in Section~\ref{sec:methods} and is available, e.g., in \cite{Chisci2001}. For nonlinear systems, the methods in \citet{Koehler2018,Koehler2020} provide suitable design mechanisms. In the following sections, we extend the predictive safety filter framework in \eqref{eq:safety_filter} such that robust asymptotic stability can additionally be guaranteed for the closed-loop system.

\section{Predictive Stability Filter}\label{sec:stab_filter}
In this section, we introduce the predictive stability filter formulation and the resulting online algorithm. 
In order to provide theoretical robust asymptotic stability guarantees of the resulting closed-loop system in Section~\ref{sec:main_result}, we require a Lyapunov function, which will be defined implicitly through the online optimization problem.
Building upon the predictive safety filter formulation in \eqref{eq:safety_filter}, the aim is then to ensure, through the online optimization, that a Lyapunov decrease is achieved at every time step. 
\par
The main modification of the predictive safety filter which is proposed is based upon ideas of suboptimal MPC, see, e.g., \cite{Scokaert1999}. Here, a Lyapunov function which is constructed through stage costs and a terminal cost of a predicted system trajectory is decreased with respect to a given warmstart input sequence through a constraint. As is standard in the MPC literature, the candidate sequence which is used to prove recursive feasibility of the optimization problem is also used to show a Lyapunov decrease of a cost function. Such a candidate sequence can therefore be used as a suitable warmstart input sequence to ensure a Lyapunov decrease. We therefore start by defining the function
\begin{equation}\label{eq:Lyapunov_function}
    V((x,\mathbf{u})) = \sum_{i=0}^{N-1}l(\phi(x,\mathbf{u};i), u_i) + m(\phi(x,\mathbf{u};N)),
\end{equation}
which consists of stage costs $l\vcentcolon\mathbb{R}^n\times\mathbb{R}^m\rightarrow \mathbb{R}$ and a terminal cost $m\vcentcolon\mathbb{R}^n\rightarrow \mathbb{R}$ and where $\phi(x,\mathbf{u};k) \vcentcolon \mathbb{R}^{n_x} \times \mathbb{R}^{n_u \cdot (k-1)}  \rightarrow \mathbb{R}^{n_x}$ denotes the $k$-steps forward time simulation of the nominal system $x^+=f(x,u,0)$ for some input sequence $\mathbf{u}$. These cost terms are subject to the following assumptions.

\begin{assum}[Continuity]\label{assum:cont}
    The functions $f$, $l$ and $m$ are continuous, $l(0,0) = m(0) = 0$ and $m$ is non-negative.
\end{assum}
\begin{assum}[Lower-bounded stage cost]\label{assum:lower_bound}
    There exists a function $\alpha_l \in \mathcal{K}_{\infty}$ such that 
    \begin{equation*}
        l(x,u) \geq \alpha_l(\|(x,u)\|),
    \end{equation*}
    for all $(x,u) \in \bigcup_{i=0}^{N-1}\Omega_i$. 
\end{assum}

We then define the proposed \textit{stability filter} as a discrete-time optimal control problem of the form
 \begin{subequations}\label{eq:stability_filter}
\begin{align}
 \underset{\begin{subarray}{c}
    z_0, \dots, z_N \\
    v_0, \dots, v_{N-1}
 \end{subarray}}{\min} & J(\mathbf{z},\mathbf{v},p)\\
 \text{s.t. } \;\;\;  & z_{0} = x, \label{eq:constr1}\\
        	  & z_{i+1} = f(z_i,v_i, 0), \, i=0,\dots,N-1, \label{eq:constr2}\\
          	& (z_i, v_i) \in \Omega_i, \quad \quad \,\,\,\, i=0,\dots,N-1, \label{eq:constr3}\\
         	& z_N \in \mathcal{Z}_{f}, \label{eq:constr4}\\
  		    &V((f(x, v_0, 0), \xi_c(f(x,v_0,0), \mathbf{v}, 0))) \leq \nonumber\\ 
                &V((x, \mathbf{\tilde{u}})) -(1-\rho) \cdot l(x,\tilde{u}_0). \label{eq:Lyapunov_decrease}
 \end{align}
 \end{subequations}
Compared to the predictive safety filter in \eqref{eq:safety_filter}, we additionally enforce the constraint \eqref{eq:Lyapunov_decrease}. This Lyapunov constraint imposes that any feasible solution $\mathbf{\tilde{v}}$ admits a stage cost decrease $(1-\rho)\cdot l(x,\tilde{u}_0)$ at the \textit{next} time step for the Lyapunov function in \eqref{eq:Lyapunov_function}.
As the Lyapunov function \eqref{eq:Lyapunov_function} is a function of a state and an input sequence, we consider the difference between the function value of the current state and a warmstart sequence $(x,\mathbf{\tilde{u}})$ and the function value of the next predicted state $f(x,v_0,0)$ and a candidate input sequence at the next time step, which is generated according to the function $\xi_c \vcentcolon \mathbb{R}^{n_x} \times \mathbb{R}^{n_u \cdot N}\times \mathbb{R}^{n_w} \rightarrow \mathbb{R}^{n_u \cdot N}$ and which will be formalized in Assumption~\ref{assum:improved_candidate}. 
For $\rho\in[0,1)$, it is ensured that the solution of \eqref{eq:stability_filter}, does not perform worse than the currently available warmstart $\mathbf{\tilde{u}}$, which is generated according to a candidate generating function $\xi \vcentcolon \mathbb{R}^{n_x} {\times} \mathbb{R}^{n_u {\cdot} N}{\times} \mathbb{R}^{n_w} {\rightarrow} \mathbb{R}^{n_u {\cdot} N}$, which is formalized in \eqref{eq:switch}.
\par
We can now formulate the online algorithm of the proposed stability filter in Algorithm~\ref{alg:online}. 
At every time step, given the current value of $p$, $x$ and the warmstart $\mathbf{\tilde{u}}$, which is generated through the warmstart generating function $\xi$, a feasible input sequence $\mathbf{\tilde{v}}$ satisfying \cref{eq:constr1,eq:constr2,eq:constr3,eq:constr4,eq:Lyapunov_decrease} is computed online. Finally, the first input $\tilde{v}_0$ of $\mathbf{\tilde{v}}$ is applied to the system. Note that we do not require an optimal solution $\mathbf{v}^*$ of \eqref{eq:stability_filter} to be available, but only a feasible one. 

\begin{rem}
        In \eqref{eq:Lyapunov_decrease}, a decrease function of $V((x,\mathbf{\tilde{u}}))$,
    \begin{equation*}\label{eq:filterinterpolation}
    R(x,u,\rho) \vcentcolon = 
    \begin{cases}
    \begin{aligned}
    &-(1-\rho) \cdot l(x,u), \, &&\text{if} \quad \rho \in [0,1),\\
    &\rho, \quad \quad \quad \quad \quad\quad\quad\,\, &&\text{if} \quad \rho \geq 1,\\
    \end{aligned}
\end{cases}
\end{equation*}
could instead be used, in order to "interpolate" between a stable behavior of the closed-loop for $\rho\in[0,1)$ and the standard predictive safety filter \eqref{eq:safety_filter}, which is recovered by setting $\rho\rightarrow\infty$.
\end{rem}
\begin{algorithm} 
    \caption{Stability filter online algorithm}
    \begin{algorithmic}[1]\label{alg:online}
    \STATE Measure initial state $x$
    \STATE Compute initial feasible candidate sequence $\mathbf{\tilde{u}}$
    \WHILE{True}
        \STATE Obtain parameter $p$ (e.g. a performance input $u_L$ as in \cite{Wabersich2018})
        \STATE Minimize \eqref{eq:stability_filter} and obtain $\mathbf{\tilde{v}}$
        \STATE Apply $u=\tilde{v}_0$
        \STATE Measure state $x^+$ and incurred disturbance $w$
        \STATE Generate candidate sequence $\mathbf{\tilde{u}}=\xi(x^+,\mathbf{\tilde{v}},w)$
        \STATE Set $x=x^+$
    \ENDWHILE
    \end{algorithmic}
\end{algorithm}
\section{Main theoretical result}\label{sec:main_result}
In this section, we analyse the closed-loop behavior of the system under application of the proposed stability filter in Algorithm~\ref{alg:online} and show robust asymptotic stability of an extended state-warmstart system. The results build upon the analysis in \cite{Allan2017}, where we additionally ensure robust stability in the admissible set of \eqref{eq:stability_filter} \textit{by design}, rather than considering inherent robustness properties. First, we formalize a set of definitions and assumptions that lead to robust asymptotic stability. We start by defining the admissible set of \eqref{eq:stability_filter}, i.e., the values of the parameters $x$ and $\mathbf{\tilde{u}}$ such that a feasible solution to the underlying nonconvex program exists.
\begin{defn}[Set of admissible state-warmstart pairs]~ \\
We denote the set of admissible parameters $x$ and $\mathbf{\tilde{u}}$ for some fixed value $\rho\in[0,1)$ such that a feasible solution to the optimization problem \eqref{eq:stability_filter} exists as
\begin{equation}
\begin{split}\label{eq:admissible}
\bar{\mathcal{F}}_\rho {=} \{(x,\mathbf{\tilde{u}}) \mid & \;\exists \mathbf{v}, \mathbf{z} \textup{ satisfying }  \cref{eq:constr1,eq:constr2,eq:constr3,eq:constr4,eq:Lyapunov_decrease} \\
& \wedge  \exists \mathbf{\tilde{z}} \textup{ s.t. }\mathbf{\tilde{u}}, \mathbf{\tilde{z}} \textup{ satisfy } \cref{eq:constr1,eq:constr2,eq:constr3,eq:constr4}\}.
\end{split}
\end{equation}
\end{defn}
Note that along with the typical requirement for admissibility, i.e., the existence of a feasible solution to \eqref{eq:stability_filter} for a given state $x$, we additionally require existence of $\mathbf{\tilde{z}}$, such that the joint state and input constraints are satisfied for the given warmstart input sequence $\mathbf{\tilde{u}}$ at the state $x$.
Similarly, we define a set that contains all the possible input sequences for a fixed parameter value $x$ and $\mathbf{\tilde{u}}$.
\begin{defn}[Set of feasible input sequences] Given a \\ fixed value of $\rho \in [0,1)$, let $\mathcal{U}_{\rho}(x,\mathbf{\tilde{u}}) \subseteq \mathbb{R}^{N\cdot n_u}$ denote the set of feasible input sequences for a given 
    value of the parameters $(x,\mathbf{\tilde{u}})\in\bar{\mathcal{F}}_\rho$, i.e.,
    \begin{equation*}
        \begin{aligned}
    &&\mathcal{U}_{\rho}(x,\mathbf{\tilde{u}}) \vcentcolon = &\{ \mathbf{v} \in \mathbb{R}^{N \cdot n_u}\,  \mid 
    \exists \mathbf{z} \textup{ s.t. } \mathbf{v}, \mathbf{z} \textup{ satisfy }\cref{eq:constr1,eq:constr2,eq:constr3,eq:constr4,eq:Lyapunov_decrease}\}.
    \end{aligned}
    \end{equation*}
\end{defn}
In order to establish stability of the closed-loop system for a given $\rho\in[0,1)$, we refer to an \textit{extended state} $s = (x,\mathbf{\tilde{u}})$ as, e.g., in \cite{Allan2017}, and define the difference inclusion
\begin{align}\label{eq:diffinc}
    \begin{split}
    s^+ \!\in\! H_{\rho}(s,w) \!\vcentcolon=\! \{ (x^+, \mathbf{\tilde{u}^+}) \mid & x^+ = f(x,\tilde{v}_0,w), \\
    &\mathbf{\tilde{u}^+} = \xi(f(x, \tilde{v}_0, w), \mathbf{\tilde{v}}, w), \\
    &\mathbf{\tilde{v}} \in \mathcal{U}_{\rho}(x,\mathbf{\tilde{u}})\},
    \end{split}
\end{align}
which, for a given state $x$, warmstart $\mathbf{\tilde{u}}$ and disturbance $w$ 
captures the evolution of the augmented dynamics for \textit{any} feasible control sequence $\mathbf{\tilde{v}} \in \mathcal{U}_{\rho}(x,\mathbf{\tilde{u}})$. 
\par
    The description of the coupled dynamics as a difference inclusion allows us to treat the evolution of both the warmstart $\mathbf{\tilde{u}}$ and the state $x$ of the system to be controlled for any feasible 
    solution to \eqref{eq:stability_filter} and any value of the parameter $p$, which typically changes from one instance to another. This is possible since $p$ enters the cost only, and hence does not affect the feasible set, but only the optimal solution.
    In this way, we can avoid restricting our attention to an a-priori prescribed sequence of parameter values, which is typically not known in practice. At the same time, we obtain a more general result that does not require the computation of the optimal solution to \eqref{eq:stability_filter}, but only of a feasible solution.
\par
We use the following definition, which formalizes the notion of robust asymptotic stability for difference inclusions and where we denote an arbitrary solution of the difference inclusion \eqref{eq:diffinc} after $k$ time steps as $\psi(s,\mathbf{w};k)$ for some extended state $s$ and disturbance signal $\mathbf{w}$.
\begin{defn}[Robust asymptotic stability]
    Let $\mathcal{Y}$ be a closed robust positive invariant set for the difference inclusion $s^+ \in H_{\rho}(s,w)$ for $w\in\mathcal{W}$. We say that the origin is robustly asymptotically stable with region of attraction $\mathcal{Y}$ for $s^+ \in H_{\rho}(s,w)$ if there exist $\beta \in \mathcal{KL}$ and $\gamma \in \mathcal{K}$, such that for any $s \in \mathcal{Y}$, $w_i\in\mathcal{W}$ with $0\leq i \leq k$ and for all $k \in \mathbb{N}$, any solution $\psi(s,\mathbf{w};k)$ satisfies
    \begin{equation*}
        \| \psi(s,\mathbf{w};k)\|\leq \beta(\| s\|,k) + \gamma (\| \mathbf{w} \|).
    \end{equation*}
\end{defn}
Note that compared to \cite{Allan2017}, we consider robust asymptotic stability on a robust positive invariant set for a prior known disturbance set $\mathcal{W}$. As typically done with standard difference equations, we link asymptotic stability of a difference inclusion to the existence of a Lyapunov function defined as follows.
\begin{defn}[Input-to-state Lyapunov function]\label{defn:lyapunov}~ \\
    The function $V$ is an input-to-state (ISS) Lyapunov function in the robust positive invariant set $\mathcal{Y}$ for the difference inclusion 
    $s^+ \in H_{\rho}(s,w)$ with $w\in\mathcal{W}$, if there exist 
    functions $\alpha_1, \alpha_2, \alpha_3 \in \mathcal{K}_{\infty}$, and a function $\sigma \in \mathcal{K}$ such that, for
    all $s \in \mathcal{Y}$ and for all $w\in\mathcal{W}$, 
    \begin{subequations}
    \begin{align}
        \alpha_1(\| s\|) \leq &V(s) \leq \alpha_2( \|s\|), \label{eq:Lyap_bound}\\
        \underset{\begin{subarray}{c}
    s^+ \in H_{\rho}(s,w)
 \end{subarray}}{\sup} V(s^+) \leq &V(s) - \alpha_3(\|s\|) + \sigma(\|w\|). \label{eq:Lyap_cond}
    \end{align}
    \end{subequations}
\end{defn}
Note that, compared to the standard definition of an ISS Lyapunov function as, e.g., in \cite[Definition 17]{Allan2017}, we impose the decrease condition over all the possible realizations of the disturbance $w\in\mathcal{W}$. Such a definition is consistent with definitions of ISS Lyapunov functions for difference equations in, e.g., \cite[Definition B.46]{Rawlings2017} and \cite[Definition 7]{Limon2009}.
\par
Given the definitions of robust asymptotic stability and ISS Lyapunov functions, we can use the following proposition from \cite[Lemma B.47]{Rawlings2017} to establish the sufficient relationship between ISS Lyapunov functions and robust asymptotic stability of difference inclusions. Although the proposition is stated for a difference inclusion rather than a difference equation in \cite[Lemma B.47]{Rawlings2017}, the difference is minor due to the fact that the supremum is considered in the cost decrease in \eqref{eq:Lyap_cond}.
\begin{prop}\label{prop:Lyap_stab} If the difference inclusion $s^+ \in H_{\rho}(s,w)$ admits a continuous ISS Lyapunov function in a robust positive invariant set $\mathcal{Y}$ for all $w\in\mathcal{W}$,
    then the origin is robustly asymptotically stable in $\mathcal{Y}$ for all $w\in\mathcal{W}$.
\end{prop}
In the following, we state the required assumptions and definitions allowing to show that $V((x,\mathbf{u}))$ in \eqref{eq:Lyapunov_function} defines an ISS Lyapunov function according to Definition~\ref{defn:lyapunov}, implying robust asymptotic stability of the resulting closed-loop extended state. 
In order to ensure that $\| \mathbf{\tilde{u}} \|\rightarrow 0$ as $\| x \|\rightarrow 0$, we require two distinct candidate generating functions, i.e. $\xi_c$, which is used in \eqref{eq:Lyapunov_decrease} and $\xi_f$.
The warmstart input sequence $\mathbf{\tilde{u}}$, which is used in \eqref{eq:stability_filter} is then generated given a feasible solution in the last time step using $\xi_c$ and switches to $\xi_f$ whenever $x \in \mathcal{Z}_f$ and the cost is reduced, i.e. 
\begin{equation}\label{eq:switch_candidate}
    \begin{aligned}
    V((x, \xi_f(x, \mathbf{u}))) \leq V((x, \xi_c(x, \mathbf{u}, w)))
    \end{aligned}
\end{equation}
holds. The warmstart generating function is therefore given by
\begin{equation}\label{eq:switch}
    \xi(x,\mathbf{u},w) \vcentcolon = 
    \begin{cases}
        \xi_f(x,\mathbf{u}), \, \text{if} \, x \in \mathcal{Z}_f\,\text{and}\, \eqref{eq:switch_candidate}\\
        \xi_c(x,\mathbf{u},w), \, \text{else}.
    \end{cases}
\end{equation}
The assumptions on the predictive safety filter design and the candidate generating functions are then given as follows.
\begin{assum}\label{assum:improved_candidate}
    We assume that:
    \begin{enumerate}
\item (Set design). The set $\mathcal{Z}_f$ is compact and contains the origin in its interior and the sets $\Omega_i\subseteq\mathcal{Z}$ are compact.
\item (Recursive feasibility of the candidate generating function). 
    The set $\bar{\mathcal{F}}_\rho$
     is robustly positive invariant for any fixed $\rho\in[0,1)$ for the dynamical system
     \begin{align*}
            x^+ &= f(x, u_0, w), \\
            \mathbf{u}^+ &= \xi_c(f(x, u_0, w), \mathbf{u}, w),
         \end{align*}
     for all $w \in \mathcal{W}$ and $\xi_c$ is continuous. 
\item  (Candidate nominal improvement). It holds that
\begin{equation*}
    \begin{aligned}
    && &V((f(x, u_0, 0), \xi_c(f(x, u_0, 0), \mathbf{u}, 0)))  \\
    && \leq &V((x,\mathbf{u})) - \alpha_l(\| (x, u_0)\|),    
\end{aligned}
    \end{equation*}
    for all $(x,\mathbf{u}) \in \bar{\mathcal{F}}_\rho$.
\item (Recursive feasibility of the second candidate generating function).
    There exists a set $\bar{\mathcal{F}}\subseteq\bar{\mathcal{F}}_\rho$ for all $\rho\in[0,1)$ such that $\bar{\mathcal{F}}$ is robustly positively invariant for the dynamical system
    \begin{align*}
            x^+ &= f(x, u_0, w), \\
            \mathbf{u}^+ &= \xi_f(f(x, u_0, w), \mathbf{u}),
    \end{align*}
    for any $w\in\mathcal{W}$ and any $x\in\mathcal{Z}_f$. 
\item (Terminal cost)  The candidate generating function $\xi_f$ and terminal cost $m$ fulfill
\begin{equation*}
    V((x, \xi_f(x,\mathbf{u}))) \leq m(x) \; \forall x \in \mathcal{Z}_f.
\end{equation*}
\end{enumerate}
\end{assum}
We note that the design requirements in Assumption~\ref{assum:improved_candidate} are standard in many robust MPC methods. Recursive feasibility with respect to a given disturbance set is typically satisfied by considering the error propagation with respect to a nominal trajectory. A suitable candidate sequence is then achieved by shifting the previous feasible solution while additionally accounting for the encountered disturbance using a feedback control law. Such methods are available in, e.g., \cite{Chisci2001}, for linear systems and in \citet{Koehler2018,Koehler2020} for nonlinear systems. The set $\bar{\mathcal{F}}_\rho$ can therefore be seen as a lifting of the set of the admissible states, which is implicitly defined through the MPC problem, into the space of admissible states and warmstart input sequences. For any state which is admissible for the MPC problem there exists at least one corresponding feasible input trajectory, forming an element in the set $\bar{\mathcal{F}}_\rho$.
From the MPC design, a control law is typically available, which guarantees robust positive invariance of a set $\mathcal{X}_f\supseteq\mathcal{Z}_f$ and induces a local Lyapunov function $m(x)$. Such a control law can then, e.g., be applied $N$ times to obtain $\xi_f$. The set $\bar{\mathcal{F}}$ is then again given by a lifting of $\mathcal{X}_f$ into the space of states and warmstart input sequences.
A suitable linear design method is discussed in Section~\ref{sec:methods}.
\par
The following two propositions are shown in a similar form in \cite{Allan2017} and are provided for the sake of completeness. First, we show that $V$ can be lower- and upper-bounded according to Definition \ref{defn:lyapunov}.
\begin{prop}\label{prop:Lyapunov_bounds}
    Let Assumptions~\ref{assum:cont},~\ref{assum:lower_bound}~and~\ref{assum:improved_candidate} hold.
    Then there exist functions $\alpha_1, \alpha_2 \in \mathcal{K}_{\infty}$ such that for any $s \in \bar{\mathcal{F}}_\rho$
    \begin{equation}
        \alpha_1(\| s\|) \leq V(s) \leq \alpha_2(\|s\|).
    \end{equation}
    \begin{pf} 
        The proof follows from the proof of \cite[Proposition 7 and 22]{Allan2017}.
        From \eqref{eq:Lyapunov_function} and Assumptions~\ref{assum:cont} and \ref{assum:lower_bound}, it follows that 
        \begin{align*}
            V((x,\mathbf{u})) 
            & \geq \sum_{k=0}^{N-1} \alpha_l(\|(\phi(x,\mathbf{u};k), u_k )\|) \\
            & \geq \alpha_l\left(\frac{1}{N} \sum_{k=0}^{N-1} \|(\phi(x,\mathbf{u};k), u_k)\|\right) \\
            & \geq \alpha_l\left(\frac{\|(x,\mathbf{u})\|}{N}\right) := \alpha_1(\|(x,\mathbf{u})\|),
        \end{align*}
        where the second and third inequality is shown in \cite[Proposition 22]{Allan2017}.
        From the fact that the sets $\Omega_i$ and $\mathcal{Z}_f$ are compact as well as the fact $f$ and $V$ are continuous according to Assumption~\ref{assum:cont} and the definition in \eqref{eq:Lyapunov_function}, it follows that $\bar{\mathcal{F}}_\rho$ is compact for any $\rho\in [0,1)$.
        Therefore, from \cite[Proposition 7]{Allan2017} and the fact that $\bar{\mathcal{F}}_\rho$ contains the origin, it follows that there exists $\alpha_2\in\mathcal{K}_\infty$ such that $V((x,\mathbf{u}))\leq \alpha_2(\|(x,\mathbf{u})\|)$.$\hfill \qed$
    \end{pf}
\end{prop}
\par
Moreover, we will make use of the following intermediate result which allows us to upper bound the norm of $\mathbf{\tilde{u}}$ with a function of $x$ and which is necessary in order to show that $\| x\| \to 0$ implies $\| \mathbf{\tilde{u}} \| \to 0$. 
\begin{prop}\label{prop:u_upper_bound}
    Let Assumptions~\ref{assum:cont},~\ref{assum:lower_bound}~and~\ref{assum:improved_candidate} hold.
    Then there exists a function $\alpha_c \in \mathcal{K}_{\infty}$ such that $\| \mathbf{\tilde{u}} \| \leq \alpha_c(\| x \|)$, for any $(x, \mathbf{\tilde{u}}) \in \bar{\mathcal{F}}_\rho$.
    \begin{pf}
        This proof follows similarly to the proof of \cite[Proposition 10]{Allan2017}. For any $x\in\mathcal{Z}_f$, we have 
        \begin{align*}
            \alpha_1(\|\mathbf{\tilde{u}}\|) \leq \alpha_1(\|(x,\mathbf{\tilde{u}})\|)\leq V((x,\mathbf{\tilde{u}})) \leq m(x) \leq \alpha_m(\|x\|), 
        \end{align*}
        where $V((x,\mathbf{\tilde{u}}))\leq m(x)$ follows from Assumption~\ref{assum:improved_candidate} and the definition of $\xi$ in \eqref{eq:switch_candidate} and the upper bound $\alpha_m\in\mathcal{K}_\infty$ exists according to \cite[Proposition 7]{Allan2017}.
        It follows that $\| \mathbf{\tilde{u}} \| \leq \alpha_1^{-1} \circ \alpha_m(\|x\|) := \hat{\alpha}_c(\| x \|) $.
        In the case that $x\notin\mathcal{Z}_f$, we define $\mu:=\max_{(x_i,u_i)\in\Omega_i}\| \mathbf{u} \| < \infty$ due to compactness of $\Omega_i$. As $\mathcal{Z}_f$ contains the origin in its interior according to Assumption~\ref{assum:improved_candidate}
        , it holds that $\exists r>0$ such that $\mathcal{B}(0,r){\subset}\mathcal{Z}_f$.
        By defining $\gamma:=\min\{1,\hat{\alpha}_c(r)\}$, we have $\| \mathbf{\tilde{u}} \|\leq \mu \leq (\frac{\mu}{\gamma})\hat{\alpha}_c(\|x\|) := \alpha_c(\| x \|)$ $\forall x\in\bar{\mathcal{F}}_\rho$. $\hfill \qed$
    \end{pf}
\end{prop}
We now state the main robust asymptotic stability result.
\begin{thm}[Robust asymptotic stability]\label{thm:main_result} Let Assumptions~\ref{assum:cont},~\ref{assum:lower_bound}~and~\ref{assum:improved_candidate} hold. Then, for any fixed $\rho \in [0,1)$, it holds that $\bar{\mathcal{F}}_\rho$ is robustly positive invariant, i.e. \eqref{eq:stability_filter} is recursively feasible for all $w\in\mathcal{W}$ and the origin $(x,\mathbf{\tilde{u}}) = 0$ is
    robustly asymptotically stable for all $w\in\mathcal{W}$ for the difference inclusion $s^+ \in H_{\rho}(s,w)$ with region of attraction $\bar{\mathcal{F}}_\rho$ under application of Algorithm~\ref{alg:online}.
    \begin{pf}
        We first show robust positive invariance of $\bar{\mathcal{F}}_\rho$ for $s^+ \in H_{\rho}(s,w)$ for any $\rho \in [0,1)$. 
        Given that $(x,\mathbf{\tilde{u}}) \in \bar{\mathcal{F}}_\rho$, the optimizer selects any feasible input sequence $\mathbf{\tilde{v}} \in \mathcal{U}_{\rho}(x,\mathbf{\tilde{u}})$, where $(x,\mathbf{\tilde{v}}) \in \bar{\mathcal{F}}_\rho$ trivially holds since $\mathbf{\tilde{v}} \in \mathcal{U}_{\rho}(x,\mathbf{\tilde{v}})$ for any 
        $\rho \in [0,1)$. 
        Hence using $\mathbf{\tilde{u}}^+=\xi(f(x,\tilde{v}_0,w),\mathbf{\tilde{v}},w))$, we have that $(x^+, \mathbf{\tilde{u}}^+) \in \bar{\mathcal{F}}_\rho$ for any $w \in \mathcal{W}$ according to Assumption~\ref{assum:improved_candidate}, which shows that $\bar{\mathcal{F}}_\rho$ is robustly positive invariant for $s^+ \in H_{\rho}(s,w)$.
        \par
        Subsequently, we show robust asymptotic stability of the closed-loop system by showing that $V(s)$ is a Lyapunov function in $\bar{\mathcal{F}}_\rho$ according to Definition~\ref{defn:lyapunov}. This is sufficient for robust asymptotic stability as stated in Proposition~\ref{prop:Lyap_stab}. The lower and upper bounds of $V$ in \eqref{eq:Lyap_bound}, i.e. $\alpha_1$ and $\alpha_2\in\mathcal{K}_\infty$, exist according to Proposition~\ref{prop:u_upper_bound}. 
        \par
        It now suffices to show that there exists a function $\alpha_{3} \!\in\! \mathcal{K}_{\infty}$ and a function $\sigma\in\mathcal{K}$, such that for all $s \in \bar{\mathcal{F}}_\rho$ and $w\in\mathcal{W}$ the decrease \eqref{eq:Lyap_cond} is achieved.
    We have for all $w\in\mathcal{W}$, that
    \begin{equation*}
        \begin{aligned}
        &\!\!\!\sup_{s^+ \in H_{\rho}(s,w)} V(s^+) \\
        = \;&\!\!\!\sup_{\mathbf{\tilde{v}} \in \mathcal{U}_{\rho}(x, \mathbf{\tilde{u}})} V((f(x, \tilde{v}_0, w), \xi(f(x, \tilde{v}_0, w), \mathbf{\tilde{v}}, w))) 
        \end{aligned}
    \end{equation*}
    due to the definition of the difference inclusion \eqref{eq:diffinc}. We distinguish between the two cases of the definition of $\xi$: 
    \textit{Case I.} Either $f(x, \tilde{v}_0, w) \notin \mathcal{Z}_f$ or 
    \begin{flalign*}
    f(x,\tilde{v}_0, w) \in \mathcal{Z}_f \; \wedge 
     & && V((f(x,\tilde{v}_0, w), \xi_f(f(x, \tilde{v}_0, w), \mathbf{\tilde{v}}))) \\
     &>\!\!\!\!\!&&V((f(x, \tilde{v}_0, w), \xi_c(f(x, \tilde{v}_0, w), \mathbf{\tilde{v}}, w)))\!\vcentcolon &&&
    \end{flalign*}
    In this case, it holds that 
    \begin{equation*}
        \begin{aligned}
        & V((f(x, \tilde{v}_0, w), \xi(f(x, \tilde{v}_0, w), \mathbf{\tilde{v}}, w))) \\
        =& V((f(x, \tilde{v}_0, w), \xi_c(f(x, \tilde{v}_0, w), \mathbf{\tilde{v}}, w))).
        \end{aligned}
    \end{equation*}
    \textit{Case II.} $f(x, \tilde{v}_0, w) \in \mathcal{Z}_f$ s.t.
            \begin{flalign*}
            &V((f(x, \tilde{v}_0, w), \xi_f(f(x, \tilde{v}_0, w), \mathbf{\tilde{v}}))) \\
            \leq &V((f(x, \tilde{v}_0, w), \xi_c(f(x, \tilde{v}_0, w), \mathbf{\tilde{v}}, w))): &&&
            \end{flalign*}
            In this case, it holds that
            \begin{equation*}
            \begin{aligned}
            & V((f(x, \tilde{v}_0, w), \xi(f(x, \tilde{v}_0, w), \mathbf{\tilde{v}}, w))) \\
            = & V((f(x, \tilde{v}_0, w), \xi_f(f(x, \tilde{v}_0, w), \tilde{\mathbf{v}}))) \\
            \leq & V((f(x, \tilde{v}_0, w), \xi_c(f(x, \tilde{v}_0, w), \mathbf{\tilde{v}}, w))) .
            \end{aligned}
            \end{equation*}
    We therefore have in both cases that 
    \begin{align*} 
    &\sup_{s^+ \in H_{\rho}(s,w)} V(s^+) \\
    \leq  \sup_{\mathbf{\tilde{v}} \in \mathcal{U}_{\rho}(x, \mathbf{\tilde{u}})}&V((f(x, \tilde{v}_0, w), \xi_c(f(x, \tilde{v}_0, w), \mathbf{\tilde{v}}, w)))
    \end{align*}
    Next, using compactness of $\bar{\mathcal{F}}_\rho$, as shown in the proof of Proposition~\ref{prop:Lyapunov_bounds}, and continuity of $l$, $m$, $\xi_c$ and $f$ according to Assumptions~\ref{assum:cont}~and~\ref{assum:improved_candidate}, it follows from the Heine-Cantor theorem, see, e.g. \cite[Theorem 4.19]{Rudin1953}, that $V$, $f$ and $\xi_c$ are uniformly continuous on $\bar{\mathcal{F}}_\rho$. Therefore there exist class $\mathcal{K}_\infty$ functions $\sigma_\cdot$, see, e.g., \cite[Lemma 1]{Limon2009}, such that 
    \begin{alignat*}{3}
    &\sup_{\mathbf{\tilde{v}} \in \mathcal{U}_{\rho}(x, \mathbf{\tilde{u}})}&& V((f(x, \tilde{v}_0, w), \xi_c(f(x, \tilde{v}_0, w), \mathbf{\tilde{v}}, w))) \\
    \leq & \sup_{\mathbf{\tilde{v}} \in \mathcal{U}_{\rho}(x, \mathbf{\tilde{u}})}&&V((f(x, \tilde{v}_0, 0), \xi_c(f(x, \tilde{v}_0, 0), \mathbf{\tilde{v}}, 0))) \\
      & \qquad\quad+ &&\sigma_{V,x} \circ \sigma_{f,w}(\| w \|) + \sigma_{V,u} \circ \sigma_{\xi_c,x} \circ \sigma_{f,w}(\| w \|) \\
     & \qquad\quad+ &&\sigma_{V,u} \circ \sigma_{\xi_c,w}(\| w \|) \\
    \leq & \;V((x,\mathbf{\tilde{u}}))\;-&&\; (1-\rho)\alpha_l(\| (x, \tilde{u}_0)\|) + \sigma_{V,w}(\| w \|),
    \end{alignat*}
    for any $w \in \mathcal{W}$ and any $(x, \mathbf{\tilde{u}}) \in \bar{\mathcal{F}}_\rho$, where in the last inequality we used \eqref{eq:stability_filter} and introduced the class $\mathcal{K}_\infty$ function 
    \begin{equation*}
    \sigma_{V,w}=\sigma_{V,x} \circ \sigma_{f,w} +  \sigma_{V,u} \circ \sigma_{\xi_c,x} \circ \sigma_{f,w} + \sigma_{V,u} \circ \sigma_{\xi_c,w}.
    \end{equation*}    
        In order to link the decrease in $\| (x,\tilde{u}_0) \|$ to the one in $\| (x,\mathbf{\tilde{u}})\|$, we make the following considerations:
        \begin{equation*}
            \begin{aligned}
            && \| (x, \mathbf{\tilde{u}}) \| &\leq \| x \| + \| \mathbf{\tilde{u}} \| 
            \leq \| x \| + \alpha_c(\| x\|) =\vcentcolon \alpha_{c^*}(\| x\|) \\
            && &\leq \alpha_{c^*}(\| (x,\tilde{u}_0)\|),
            \end{aligned}
        \end{equation*}  
        where we used Proposition~\ref{prop:u_upper_bound}, such that we can write 
        \begin{equation*}
            (1{-}\rho)\alpha_l\left(\| (x, \tilde{u}_0)\|\right) {\geq} (1{-}\rho)\alpha_l \circ \alpha_{c^*}^{-1}\left(\|(x,\mathbf{\tilde{u}})\| \right) {=\vcentcolon}\alpha_{3}(\| x, \mathbf{\tilde{u}}\|).
        \end{equation*}
        Therefore, we have $\forall w\in\mathcal{W}$ the desired Lyapunov decrease 
        \begin{equation*}
            \begin{aligned}
            \sup_{s^+ {\in} H_{\rho}(s,w)} V(s^+) 
            \leq V((x,\mathbf{\tilde{u}})) - \alpha_{3}(\| (x,\mathbf{\tilde{u}}) \|) + \sigma_{V,w} ( \| w \|).
            \end{aligned}
        \end{equation*}
        We have thus shown that $V$ is a valid ISS Lyapunov function according to Definition \ref{defn:lyapunov}. $\hfill \qed$
        \par
    \end{pf}
\end{thm}
\par
Note that the main difference of the proposed theoretical analysis compared to \cite{Allan2017}, is the fact that robust asymptotic stability is guaranteed \textit{by design}. Rather than relying on an inherent robustness approach, where there exists some $\delta$, such that for all $\|\mathbf{w}\|<\delta$ the origin is robustly asymptotically stable for a sublevel set of a terminal Lyapunov function, we show that the origin is robustly asymptotically stable for the set of admissible state-warmstart pairs $\bar{\mathcal{F}}_\rho$ for all $w\in\mathcal{W}$ for an \textit{a priori} given set $\mathcal{W}$. Therefore, given a warmstart input sequence which satisfies the constraints in \eqref{eq:safety_filter}, it holds that if the initial state $x$ admits a feasible solution in \eqref{eq:stability_filter}, given Assumptions~\ref{assum:cont},~\ref{assum:lower_bound}~and~\ref{assum:improved_candidate} are satisfied, the problem is recursively feasible and the closed-loop system will converge to a region around the origin.
\begin{rem}
    The compactness assumption on $\mathcal{Z}$ can be relaxed if $f$, $m$, $l$ and $\xi_c$ are Lipschitz continuous and $\mu=\sup_{(x_i,u_i)\in\Omega_i} \| \mathbf{u} \| < \infty$, e.g., in the case of decoupled constraints $\mathcal{Z}=\mathcal{X}\times\mathcal{U}$ with $u_i\in\mathcal{U}$ with compact $\mathcal{U}$.
\end{rem}

\section{Example and linear filter design} \label{sec:methods}
In this section, we demonstrate the proposed stability filter on a lane keeping example for automotive systems. The example is based on a linearization of the vehicle's dynamics and we therefore discuss a design procedure for linear systems affected by additive disturbances based on~\cite{Chisci2001}.
As Theorem~\ref{thm:main_result} provides a concept that supports more general problem settings, we refer to~\cite{Limon2009,Koehler2018,Koehler2020} for similar design procedures for nonlinear systems. We first present the general design procedure for linear systems with polytopic constraints and then specify the details of the numerical example.
\subsection{Design for linear systems and polytopic constraints}\label{sec:design}
We consider linear time-invariant systems of the form
\begin{equation}
    x^+ = f(x,u,w) = Ax + Bu + w, ~w\in\mathcal W
\end{equation}
with polytopic disturbance set $\mathcal W$ and polytopic state and input constraint sets $\mathcal{X}\subseteq\mathbb R^{n_x}$ and $\mathcal{U}\subset\mathbb R^{n_u}$ containing the origin. Stability is defined with respect to a quadratic stage cost function of the form $l(x,u) \vcentcolon = \frac{1}{2}x^{\top}Qx + \frac{1}{2}u^{\top}Ru$ with terminal cost $m(x) \vcentcolon = \frac{1}{2}x^{\top}Px$ and symmetric and positive definite matrices $Q, R$ and $P$ as commonly used in model predictive control literature~\cite{Rawlings2017}.
%
%
The general stability filter~\eqref{eq:stability_filter} in the linear setting is given by
\begin{equation}\label{eq:lqmpc_sf}
\begin{aligned}
&\underset{\begin{subarray}{c}
    z_0, \dots, z_N \\
    v_0, \dots, v_{N-1}
\end{subarray}}{\min}	
&&J(\mathbf{z},\mathbf{v},p)\\
&\quad \,\,\, \text{s.t.}   && z_{0} = x, \\
&\quad          		   	&& z_{i+1} = Az_i + Bv_i, \,\, i=0,\dots,N-1, \\
&\quad          		   	&& v_i \in \mathcal{U}\ominus K\mathcal{E}_i, \quad \quad\hspace{0.15em} i=0,\dots,N-1, \\
&\quad          		   	&& z_i \in \mathcal{X}\ominus\mathcal{E}_i, \quad\quad \quad i=0,\dots,N-1, \\
&\quad          		   	&& z_N \in \mathcal{X}_{f}\ominus \mathcal{E}_N, \\
&\quad  		   	        &&V((Ax + Bv_0, \xi_c(Ax + Bv_0, \mathbf{v}, 0))) \leq \\
&\quad                      &&V((x, \tilde{\mathbf{u}})) - (1-\rho) \cdot l(x, \tilde u_0). \\
\end{aligned}
\end{equation}
The constraint tightening is based on a stabilizing disturbance feedback, which is given by $u^+=v^+ +K(x^+-z^+)=v^+ + Kw$, where $K \in \mathbb{R}^{n_x \times n_u}$ is constant. By iteratively applying this control law, we obtain the so-called candidate solution, which is used to establish recursive feasibility in~\cite{Chisci2001} and serves as a candidate generating function for the stability filter~\eqref{eq:lqmpc_sf}:
\begin{align}\label{eq:lti_tube_xi_c}
        \xi_c(x,\mathbf{v},w)\vcentcolon=
        (&v_1 + Kw, \; v_2+ K(A+BK)^{1}w, \\
        &\dots, K\phi(x,\mathbf{v};N) + K(A+BK)^{N-1}w). \nonumber
\end{align}
The state and input tightenings $\mathcal{E}_i$ and $K\mathcal{E}_i$ are chosen to constrain the nominal predicted states and inputs such that application of~\eqref{eq:lti_tube_xi_c} yields a feasible candidate solution for~\eqref{eq:lqmpc_sf} at $x^+$ for all perturbations $w\in\mathcal W$. More precisely, all possible propagations of the perturbation $w=(x^+-z^+)$ through the Minkowski sum $\mathcal{E}_i \vcentcolon = \bigoplus_{j=0}^{i-1}(A+BK)^j\mathcal{W}$ with corresponding error feedback $K\mathcal{E}_i$ are considered.
The last predicted state is required to be contained in a tightened terminal set $\mathcal{Z}_f=\mathcal X_f\ominus\mathcal E_N$ to ensure that it will be reached under application of~\eqref{eq:lti_tube_xi_c} for all $w\in\mathcal W$.
Once the terminal set is reached, we switch to a stabilizing controller of the form $u=Kx$, as it can be seen from the last entry of~\eqref{eq:lti_tube_xi_c}. Thereby, the set $\mathcal X_f$ is computed to satisfy the following assumption.
\begin{assum}\label{assum:rpi_chisci}
    The compact set $\mathcal{X}_f\!\!\subseteq\!\!\mathcal {X}$ contains the origin and is robust positive invariant, i.e., $(A {+} BK)x \!+\! w \!\in\! \mathcal{X}_f$ $\forall x \!\in\!\mathcal{X}_f$, $w \!\in\! \mathcal{W}$, and it holds that $\forall x\!\in\!\mathcal{X}_f$, $Kx\!\in\!\mathcal{U}$.
\end{assum}
By selecting the terminal candidate generating function as
\begin{equation}
    \xi_f(x, \mathbf{v}) \vcentcolon = (Kx, K \phi(x,Kx;1), \dots),
\end{equation}
we then obtain $P$ by solving
\begin{equation}\label{eq:lyap_decrease_chisci}
    \frac{1}{2}x^{\top}(A + BK)^{\top}P(A + BK)x - \frac{1}{2}x^{\top}Px = - \frac{1}{2}x^{\top}Qx.
\end{equation}
%
\begin{cor}
    If Assumption~\ref{assum:rpi_chisci} holds, then it follows that the stability filter~\eqref{eq:lqmpc_sf} satisfies all assumptions of Theorem~\ref{thm:main_result} for all $\rho\in[0,1)$.
    \begin{pf}
        We consider $\rho\!=\!0$, implying the result for all $\rho\!\in\! [0,1)$ due to the linearity of the dynamics and the generating functions.
        Assumption~\ref{assum:cont} holds
        for linear dynamics and quadratic cost terms. Assumption~\ref{assum:lower_bound} holds due to positive definiteness of $Q$. Assumptions~\ref{assum:improved_candidate}.1, \ref{assum:improved_candidate}.4 and \ref{assum:improved_candidate}.5 hold with $\mathcal{Z}_f\!=\!\mathcal{X}_f\ominus\mathcal{E}_N$ and the definition of $\mathcal{E}_i$ due to Assumption~\ref{assum:rpi_chisci} together with \eqref{eq:lyap_decrease_chisci}. Assumptions~\ref{assum:improved_candidate}.2~and~\ref{assum:improved_candidate}.3
        can be verified through the proof for recursive feasibility in \cite{Chisci2001} using $\xi_c$ and the decrease \eqref{eq:lyap_decrease_chisci}. $\hfill \qed$
    \end{pf}
\end{cor}
\subsection{Stability filter for motion control}\label{subsec:motion_control}
\begin{figure*}
    \includegraphics[width=0.95\linewidth]{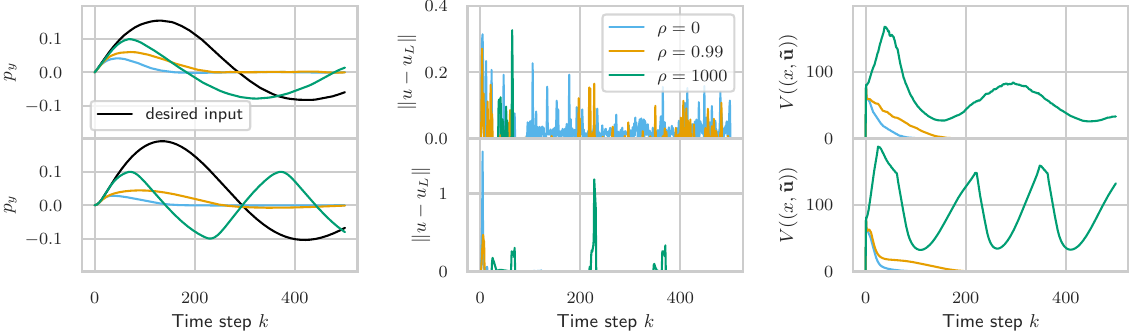}
    \caption{Numerical example of vehicle motion control according to Section~\ref{subsec:motion_control} for different values of $\rho$ in~\eqref{eq:Lyapunov_decrease} are shown. Linear (top) and nonlinear (bottom) vehicle simulations are performed with the task of recovering from a destabilizing initial state. The lateral offset with respect to a straight road is shown in the left column, where stability filters are applied to mitigate constraint violations. The middle column shows the impact of reducing the stability filter interventions by relaxing the stability filter constraint for larger values of $\rho$. The right column displays the corresponding impact on the Lyapunov function. 
    \label{fig:illustrations_experiments}}
\end{figure*}
The following numerical example demonstrates the application of the proposed method to ensure stability and constraint satisfaction for motion control.
To this end, we consider the task of safely restoring the stability of a miniature vehicle traveling along a straight road that has been destabilized by passing a low-friction surface.
For illustrative purposes, we will choose a potentially unsafe state feedback controller that models a human driver or a suboptimal learning-based controller, which we wish to apply, if safe and stabilizing, i.e., $J(\mathbf{z},\mathbf{v},p):=\Vert v_0- p\Vert$ with desired input $p=u_L$.
\par
The miniature vehicle dynamics are described using a nonlinear dynamic bicycle model with Pacejka tire forces as described in~\cite[Section IV]{tearle2021predictive}, where we scale the forces by $0.3$ to model low friction. The system has states $x\in\mathbb R^5$, $x = [p_y,\Psi,v_x,v_y,r]$ and inputs $u = [\delta, \tau]$, where $p_y$ describes the lateral offset along a straight road and $\Psi$ is the relative heading angle; $v_x$, $v_y$ and $r$ are the velocities and the yaw rate of change of the vehicle body frame. Finally, $\delta$ is the steering angle and $\tau$ is the drive command.
The continuous system dynamics are discretized using a zero-order input hold over a sampling period of $10$ milliseconds, integrated using a 4th-order Runge-Kutta method to obtain $x^+=f(x,u)$.
The safety constraints are given by a maximal lateral offset $|p_y|\leq 0.1$ and stability is defined with respect to the stage cost function $l(x,u) \vcentcolon = \frac{1}{2}x^{\top}Qx + \frac{1}{2}u^{\top}Ru^{\top}$ with $Q=4\cdot I_5$ and $R=2\cdot I_2$.
We apply the design procedure presented in Section~\ref{sec:design} by linearizing the nonlinear system at  $1.1~[\mathrm{ms}^{-1}]$. The polytopic set $\mathcal W$ from which the disturbances are drawn is chosen to approximately cover the neglected nonlinearities, i.e., $Ax+Bu-f(x,u)$, within smaller state bounds given by $-\bar x \leq x \leq \bar x$ with $\bar x = [0.1,0.26,1.3,0.5,\pi/2,0.61]$.
Selecting $K$ and $P$ as the LQR solution with respect to $Q$ and $R$ allows us to apply the design procedure described in Section~\ref{sec:design}, where we have computed the constraint tightenings by solving a linear programming problem per half-space using SciPy~\citep{huangfu2018parallelizing}. The terminal set according to Assumption~\ref{assum:rpi_chisci} has been computed as described, e.g., in \cite{borrelli2017predictive} using the polytope toolbox within the TuLiP library~\citep{filippidis2016control}.
\par
The simulations were performed for different values of $\rho$, ranging from strictly stabilizing ($\rho\!=\!0$), relaxed stabilizing ($\rho\!=\!0.99$), to the predictive safety filter case ($\rho\!=\!1000$). The results are presented in Figure~\ref{fig:illustrations_experiments}.
Simulations were performed for each value of $\rho$ using both the linearized system dynamics (Figure~\ref{fig:illustrations_experiments}, top) and the nonlinear dynamics (Figure~\ref{fig:illustrations_experiments}, bottom). The linear system simulation uses uniformly sampled additive disturbances from the set $\mathcal{W}$ for which the stability filter was designed, while in the nonlinear case the nominal dynamics are used.
Although the desired control action may result in constraint violations and fail to satisfy the desired stability properties, the application of the safety filter ($\rho\!=\!1000$) ensures constraint satisfaction. Additionally, the application of the stability filter ($\rho\!=\!0,\rho\!=\!0.99)$ enforces the stability criteria as per Theorem~\ref{thm:main_result}, even in the presence of disturbances.
Relaxing the stability criteria from $\rho\!=\!0$ to $\rho\!=\!0.99$ results in fewer interventions, while still ensuring stability with slower convergence. This provides a tuning parameter for practical applications supported by the theoretical analysis of the proposed method. We note that the stability filter designed for the linear system achieves similar performance when applied to the nominal nonlinear system dynamics. In this case, the interventions are less frequent due to the accuracy of the linearized model around the steady-state, while in the linear case, the disturbance is drawn uniformly from $\mathcal{W}$ at every time step, resulting in interventions due to the stability constraint even close to the steady-state.
\section{Conclusion}
We propose an extension to predictive safety filters, which adds stability properties by imposing a Lyapunov decrease compared to a given warmstart input sequence in an online optimization problem. Robust asymptotic stability of the resulting closed-loop system is guaranteed for the extended state-warmstart system using standard robust MPC design procedures. By providing an analysis of the resulting difference inclusion, stability of the closed-loop can be guaranteed by computing any feasible input sequence of the resulting online optimization problem. Finally, the proposed framework is illustrated on a lane keeping example, successfully stabilizing the vehicle's dynamics.
\section*{DATA AVAILABILITY STATEMENT}
The data of this study is available in the ETH Research Collection: \url{doi.org/10.3929/ethz-b-000668027}.
\bibliography{bibliography}             

\begin{thebibliography}{26}
\providecommand{\natexlab}[1]{#1}
\providecommand{\url}[1]{\texttt{#1}}
\providecommand{\urlprefix}{URL }
\expandafter\ifx\csname urlstyle\endcsname\relax
  \providecommand{\doi}[1]{doi:\discretionary{}{}{}#1}\else
  \providecommand{\doi}{doi:\discretionary{}{}{}\begingroup \urlstyle{rm}\Url}\fi

\bibitem[{Agrawal and Sreenath(2017)}]{Agrawal2017}
Agrawal, A. and Sreenath, K. (2017).
\newblock Discrete control barrier functions for safety-critical control of discrete systems with application to bipedal robot navigation.
\newblock In \emph{Robotics: Science and Systems}, volume~13, 1--10. Cambridge, MA, USA.

\bibitem[{Akametalu et~al.(2014)Akametalu, Fisac, Gillula, Kaynama, Zeilinger, and Tomlin}]{Akametalu2014}
Akametalu, A.K., Fisac, J.F., Gillula, J.H., Kaynama, S., Zeilinger, M.N., and Tomlin, C.J. (2014).
\newblock Reachability-based safe learning with {G}aussian processes.
\newblock In \emph{53rd IEEE Conference on Decision and Control}, 1424--1431.

\bibitem[{Allan et~al.(2017)Allan, Bates, Risbeck, and Rawlings}]{Allan2017}
Allan, D.A., Bates, C.N., Risbeck, M.J., and Rawlings, J.B. (2017).
\newblock On the inherent robustness of optimal and suboptimal nonlinear {MPC}.
\newblock \emph{Systems \& Control Letters}, 106, 68--78.

\bibitem[{Borrelli et~al.(2017)Borrelli, Bemporad, and Morari}]{borrelli2017predictive}
Borrelli, F., Bemporad, A., and Morari, M. (2017).
\newblock \emph{Predictive control for linear and hybrid systems}.
\newblock Cambridge University Press.

\bibitem[{Chisci et~al.(2001)Chisci, Rossiter, and Zappa}]{Chisci2001}
Chisci, L., Rossiter, J., and Zappa, G. (2001).
\newblock {S}ystems with persistent disturbances: predictive control with restricted constraints.
\newblock \emph{Automatica}, 37, 1019--1028.

\bibitem[{Didier et~al.(2023)Didier, Jacobs, Sieber, Wabersich, and Zeilinger}]{Didier2023}
Didier, A., Jacobs, R.C., Sieber, J., Wabersich, K.P., and Zeilinger, M.N. (2023).
\newblock Approximate predictive control barrier functions using neural networks: a computationally cheap and permissive safety filter.
\newblock In \emph{2023 European Control Conference (ECC)}. IEEE.

\bibitem[{Filippidis et~al.(2016)Filippidis, Dathathri, Livingston, Ozay, and Murray}]{filippidis2016control}
Filippidis, I., Dathathri, S., Livingston, S.C., Ozay, N., and Murray, R.M. (2016).
\newblock Control design for hybrid systems with tulip: The temporal logic planning toolbox.
\newblock In \emph{2016 IEEE Conference on Control Applications}, 1030--1041.

\bibitem[{Fisac et~al.(2019)Fisac, Akametalu, Zeilinger, Kaynama, Gillula, and Tomlin}]{Fisac2019}
Fisac, J.F., Akametalu, A.K., Zeilinger, M.N., Kaynama, S., Gillula, J., and Tomlin, C.J. (2019).
\newblock A general safety framework for learning-based control in uncertain robotic systems.
\newblock \emph{IEEE Transactions on Automatic Control}, 64(7), 2737--2752.

\bibitem[{Greeff et~al.(2021)Greeff, Hall, and Schoellig}]{Greeff2021}
Greeff, M., Hall, A.W., and Schoellig, A.P. (2021).
\newblock Learning a stability filter for uncertain differentially flat systems using {G}aussian processes.
\newblock In \emph{2021 60th IEEE Conference on Decision and Control (CDC)}, 789--794.

\bibitem[{Heidarinejad et~al.(2012)Heidarinejad, Liu, and Christofides}]{Heidarinejad2012}
Heidarinejad, M., Liu, J., and Christofides, P.D. (2012).
\newblock Economic model predictive control of nonlinear process systems using {L}yapunov techniques.
\newblock \emph{AIChE Journal}, 58(3), 855--870.

\bibitem[{Huangfu and Hall(2018)}]{huangfu2018parallelizing}
Huangfu, Q. and Hall, J.J. (2018).
\newblock Parallelizing the dual revised simplex method.
\newblock \emph{Mathematical Programming Computation}, 10(1), 119--142.

\bibitem[{K{\"o}hler et~al.(2018)K{\"o}hler, M{\"u}ller, and Allg{\"o}wer}]{Koehler2018}
K{\"o}hler, J., M{\"u}ller, M.A., and Allg{\"o}wer, F. (2018).
\newblock A novel constraint tightening approach for nonlinear robust model predictive control.
\newblock In \emph{2018 Annual American control conference (ACC)}, 728--734. IEEE.

\bibitem[{K{\"o}hler et~al.(2020)K{\"o}hler, Soloperto, M{\"u}ller, and Allg{\"o}wer}]{Koehler2020}
K{\"o}hler, J., Soloperto, R., M{\"u}ller, M.A., and Allg{\"o}wer, F. (2020).
\newblock A computationally efficient robust model predictive control framework for uncertain nonlinear systems.
\newblock \emph{IEEE Transactions on Automatic Control}, 66(2), 794--801.

\bibitem[{Limon et~al.(2009)Limon, Alamo, Raimondo, De~La~Pe{\~n}a, Bravo, Ferramosca, and Camacho}]{Limon2009}
Limon, D., Alamo, T., Raimondo, D.M., De~La~Pe{\~n}a, D.M., Bravo, J.M., Ferramosca, A., and Camacho, E.F. (2009).
\newblock Input-to-state stability: a unifying framework for robust model predictive control.
\newblock \emph{Nonlinear Model Predictive Control: Towards New Challenging Applications}, 1--26.

\bibitem[{McAllister and Rawlings(2023)}]{Mcallister2023}
McAllister, R.D. and Rawlings, J.B. (2023).
\newblock A suboptimal economic model predictive control algorithm for large and infrequent disturbances.
\newblock \emph{IEEE Transactions on Automatic Control}.

\bibitem[{Mhaskar et~al.(2006)Mhaskar, El-Farra, and Christofides}]{Mhaskar2006}
Mhaskar, P., El-Farra, N.H., and Christofides, P.D. (2006).
\newblock Stabilization of nonlinear systems with state and control constraints using {L}yapunov-based predictive control.
\newblock \emph{Systems \& Control Letters}, 55(8), 650--659.

\bibitem[{Pannocchia et~al.(2011)Pannocchia, Rawlings, and Wright}]{Pannocchia2011}
Pannocchia, G., Rawlings, J., and Wright, S. (2011).
\newblock Conditions under which suboptimal nonlinear {MPC} is inherently robust.
\newblock \emph{System \& Control Letters}, 60(9), 747--755.

\bibitem[{Rawlings et~al.(2017)Rawlings, Mayne, and Diehl}]{Rawlings2017}
Rawlings, J.B., Mayne, D.Q., and Diehl, M.M. (2017).
\newblock \emph{Model predictive control: theory, computation, and design}.
\newblock Nob Hill, 2nd edition.

\bibitem[{Rudin(1953)}]{Rudin1953}
Rudin, W. (1953).
\newblock \emph{Principles of mathematical analysis}.
\newblock McGraw-Hill.

\bibitem[{Scokaert and Rawlings(1999)}]{Scokaert1999}
Scokaert, P. and Rawlings, J. (1999).
\newblock Feasibility issues in linear model predictive control.
\newblock \emph{AIChE Journal}, 45(8), 1649--1659.

\bibitem[{Soloperto et~al.(2020)Soloperto, K{\"o}hler, and Allg{\"o}wer}]{Soloperto2020}
Soloperto, R., K{\"o}hler, J., and Allg{\"o}wer, F. (2020).
\newblock Augmenting {MPC} schemes with active learning: intuitive tuning and guaranteed performance.
\newblock \emph{IEEE Control Systems Letters}, 4(3), 713--718.

\bibitem[{Tearle et~al.(2021)Tearle, Wabersich, Carron, and Zeilinger}]{tearle2021predictive}
Tearle, B., Wabersich, K.P., Carron, A., and Zeilinger, M.N. (2021).
\newblock A predictive safety filter for learning-based racing control.
\newblock \emph{IEEE Robotics and Automation Letters}, 6(4), 7635--7642.

\bibitem[{Wabersich and Zeilinger(2018)}]{Wabersich2018}
Wabersich, K. and Zeilinger, M. (2018).
\newblock Linear model predictive safety certification for learning-based control.
\newblock In \emph{Proceedings of the IEEE Conference on Decision and Control (CDC)}.

\bibitem[{Wabersich et~al.(2023)Wabersich, Taylor, Choi, Sreenath, Tomlin, Ames, and Zeilinger}]{Wabersich2023}
Wabersich, K.P., Taylor, A.J., Choi, J.J., Sreenath, K., Tomlin, C.J., Ames, A.D., and Zeilinger, M.N. (2023).
\newblock Data-driven safety filters: Hamilton-{J}acobi reachability, control barrier functions, and predictive methods for uncertain systems.
\newblock \emph{IEEE Control Systems Magazine}, 43(5), 137--177.

\bibitem[{Wabersich and Zeilinger(2021)}]{Wabersich2021}
Wabersich, K.P. and Zeilinger, M.N. (2021).
\newblock A predictive safety filter for learning-based control of constrained nonlinear dynamical systems.
\newblock \emph{Automatica}, 129, 109597.

\bibitem[{Zeilinger et~al.(2014)Zeilinger, Raimondo, Domahidi, Morari, and Colin}]{Zeilinger2014}
Zeilinger, M., Raimondo, D., Domahidi, A., Morari, M., and Colin, J. (2014).
\newblock On real-time robust model predictive control.
\newblock \emph{Automatica}, 50(3), 683--694.

\end{thebibliography}

\end{document}